\documentclass[prl,%
 reprint,
 amsmath,amssymb,
 aps,
]{revtex4-2}

\usepackage{graphicx}
\usepackage{dcolumn}
\usepackage{bm}

\usepackage[utf8]{inputenc}
\usepackage{newunicodechar}
\newunicodechar{̄}{\={}}

\begin{document}

\preprint{APS/123-QED}

\title{Mass spectra of Tetraquark States from the Spinless Salpeter Equation}

\author{Pranami Baishya$^{1}$ $^{3}$\thanks{Email: your-email@example.com}}
\author{Bhaskar Jyoti Hazarika$^{2}$ $^{3}$}

\affiliation{$^{1}$Department of Physics, Gauhati University, 
Gopinath Nagar, Guwahati, Assam, India}

\affiliation{$^{2}$On deputation, North Gauhati College, Guwahati, Assam, India}
\affiliation{$^{3}$Center of Theoretical Studies, Pandu College, Guwahati 781012, India}

\begin{abstract}
The mass spectrum of tetraquark states is examined using the spinless Salpeter equation alongside the Cornell potential. Analytical wave functions are derived through a variational approach, resulting in a specific mass formula for the bound states. This study investigates the mass spectra of heavy tetraquark states within a constituent diquark-antidiquark model. We employ a spinless Salpeter equation framework, incorporating the Cornell potential to describe the interaction between the diquark and antidiquark constituents. To account for relativistic effects, we utilize first-order perturbation theory to include the leading-order kinetic energy corrections arising from the expansion of the relativistic kinetic energy operator. A variational approach is adopted to determine the ground state masses, where the optimal variational parameter ${\beta}$ is found by minimizing the system's energy and then for higher states. We systematically calculate the masses of several experimentally observed and predicted tetraquark candidates,  across a range of strong coupling constant  values. Our results demonstrate a sensitivity of the calculated tetraquark masses to, and the model provides theoretical mass predictions that are in reasonable agreement with experimental observations for these exotic hadronic states. This work contributes to a deeper understanding of the internal structure and dynamics of heavy tetraquarks.

\end{abstract}

\maketitle


\section{INTRODUCTION}

The study of exotic hadronic states has gained significant attention in recent years following the observation of several states that cannot be easily accommodated within the conventional quark--antiquark or three-quark picture \citep{2003PhRvL..91z2001C}.
Among these exotic configurations, tetraquark states composed of two quarks and two antiquarks represent an important class of multiquark systems.
Understanding the structure and mass spectrum of tetraquarks provides valuable insight into the nonperturbative regime of Quantum Chromodynamics (QCD).

The discovery of exotic hadronic states that do not fit the conventional ${q \bar{q}}$ meson or qqq baryon templates has revolutionized our understanding of Quantum Chromodynamics (QCD). Since the observation of the X(3872) by the Belle Collaboration in 2003 [1], a plethora of "XYZ" states has been identified, particularly in the heavy quark sector.More recently, the LHCb collaboration reported the observation of a narrow structure in the 
J/${\psi-pair}$ mass spectrum, the 
X(6900), which is widely interpreted as a fully charmed tetraquark state (${cc\bar{c} \bar{c}}$) \citep{2020SciBu..65.1983L}. These discoveries have prompted intense theoretical investigation into the binding mechanisms of multiquark systems.
Among the various theoretical frameworks, the diquark-antidiquark model has emerged as a prominent candidate for describing tetraquark structures. In this picture, a tetraquark is viewed as a bound state of a color-antitriplet diquark and a color-triplet antidiquark. To describe the dynamics of such a system, it is essential to account for relativistic effects, as the constituent masses and binding energies often lie in a regime where non-relativistic approximations may be insufficient. The Spinless Salpeter Equation (SSE) provides a robust relativistic generalization of the Schrödinger equation, incorporating the relativistic kinetic energy while maintaining a manageable computational framework.

In this work, we employ the Spinless Salpeter Equation to investigate the mass spectra of heavy tetraquark states. We utilize the Cornell potential, which consists of a linear confinement term and a Coulomb-like short-range term, to model the inter-diquark interaction. This potential has been highly successful in describing heavy quarkonium states \citep{1978PhRvD..17.3090E}. To solve the resulting Hamiltonian, we apply a variational approach using an exponential trial wave function.Furthermore, we treat the higher-order relativistic kinetic terms as a perturbation to the zeroth-order non-relativistic Hamiltonian, allowing us to derive first-order corrections to both the mass and the wave function.

By minimizing the total mass with respect to the variational parameter ${\beta}$, we provide predictions for the masses of the 
X(3872),
X(4700), 
X(6900) and $T_{bb\bar{b}\bar{b}}$. Our results are compared with the latest experimental data \citep{2025PhRvD.111a4015L,mutuk2021nonrelativistic,2025PhRvL.135m1902B} as well with other models and predictions to test the validity of the diquark-antidiquark approximation and the influence of relativistic corrections on the tetraquark mass scale.\\
The X(3872) is one of the earliest and most extensively studied exotic hadrons. Its well measured mass and its importance in understanding multiquark dynamics serves as a benchmark hidden-charm tetraquark for testing the predictive capability of the present model.
      The X(4700) is selected as a higher-mass hidden-charm tetraquark candidate. Studying this state enables us to investigate whether the same theoretical framework remains applicable to heavier hidden-charm configurations.
       The X(6900), reported as a fully-charmed tetraquark candidate, represents a distinctly different quark configuration in which all constituents are heavy quarks. The inclusion of this state provides an opportunity to test the present model in the fully-heavy sector.
        The fully-bottom tetraquark $T_{bb\bar{b}\bar{b}}$ is included as a representative fully-bottom system. It has attracted considerable theoretical interest because of its unique heavy-quark dynamics and the possibility of forming a deeply bound multiquark state. Investigating this system allows us to examine the predictive capability of the present model in the bottom sector and to compare our results with existing theoretical predictions.
         The selected tetraquark states therefore span different heavy-quark configurations, ranging from hidden-charm to fully-heavy systems.

The organization of the paper is as follows.
In Sec.~II, we discuss the formalism, Sec.~III presents the results and discussion while the conclusion is made in Sec.~IV. 

\section{FORMALISM}

\subsection{Theoretical Framework: Wave Function and Perturbation}
\label{sec:wavefunction}

Our analysis of tetraquark states begins with the Spinless Salpeter Equation, which provides a relativistic description of a two-body system. The Hamiltonian of the system, incorporating both kinetic and potential energy terms, is given by:
\begin{equation}
    \left[\sqrt{-\nabla^{2}+m_{1}^{2}}+\sqrt{-\nabla^{2}+m_{2}^{2}}+V(r)\right]\psi({\bf r})=M\psi({\bf r})
    \label{eq:salpeter}
\end{equation}
where $m_1$ and $m_2$ are the masses of the constituent diquarks, $M$ is the total mass of the tetraquark and $V(r)$ is the inter-diquark potential.\\

In the present work, the tetraquark is treated within an effective diquark–antidiquark framework. The parameters $m_1$ and $m_2$ denote the effective masses of the diquark and antidiquark, respectively. Their numerical values are adopted from previous relativistic studies in which the diquark masses were obtained by solving the corresponding quark–quark bound-state problem. Consequently, these effective masses already include the internal dynamics and binding effects of the constituent quarks within the diquark. Therefore, in the present work the diquark and antidiquark are treated as effective constituents, and the Spinless Salpeter equation is employed to describe their relative motion under the Cornell interaction.

\subsection{The Hamiltonian and Potential}
The potential $V(r)$ employed in this study is the Cornell potential, which effectively describes the interaction between heavy quarks:
\begin{equation}
    V(r)=-\frac{4}{3}\frac{\alpha_{s}}{r}+\sigma r+c
    \label{eq:cornell_potential}
\end{equation}
where $\alpha_s$ is the strong coupling constant, $\sigma$ is the string tension (taken as $0.1~\text{GeV}^2$)\citep{2024EPJC...84..807K}, and $c$ is a phenomenological constant (set to $-0.8~\text{GeV}$)\citep{2022EPJC...82.1081P}.

To simplify the relativistic kinetic energy terms, we utilize an expansion for small momenta:
\begin{equation}
    \sqrt{m_{i}^{2}-\nabla^{2}}\approx m_{i}+\frac{-\nabla^{2}}{2m_{i}}-\frac{\nabla^{4}}{8m_{i}^{3}}
    \label{eq:kinetic_expansion}
\end{equation}
Substituting this expansion into the Hamiltonian and defining the reduced mass $\mu = (m_1 m_2) / (m_1 + m_2)$, the full Hamiltonian can be written as:
\begin{equation}
    H=m_{1}+m_{2}+\frac{p^{2}}{2\mu}-\frac{p^{4}}{8}\left(\frac{1}{m_{1}^{3}}+\frac{1}{m_{2}^{3}}\right)+V(r)
    \label{eq:full_hamiltonian}
\end{equation}
where we have replaced $\nabla^2$ with $-p^2$ (with $\hbar=1$).

For our analysis, we treat the leading non-relativistic terms as the zeroth-order Hamiltonian $H_0$ and the higher-order relativistic correction as a perturbation $H'$.
The zeroth-order Hamiltonian, which corresponds to the non-relativistic Schrödinger equation, is:
\begin{equation}
    H_0 = m_1+m_2+\frac{p^2}{2\mu}+V(r)
    \label{eq:h0}
\end{equation}
The perturbing Hamiltonian $H'$ accounts for the first-order relativistic correction to the kinetic energy:
\begin{equation}
    H' = -\frac{p^4}{8}\left(\frac{1}{m_{1}^{3}}+\frac{1}{m_{2}^{3}}\right)
    \label{eq:hprime}
\end{equation}

\subsection{Zeroth-Order Wave Function and Normalization}
We consider the ground state wave function for the zeroth-order Hamiltonian $H_0$. For an S-wave state in spherical coordinates, we adopt a simple exponential form:
\begin{equation}
    \psi^{(0)}(r)=N_0e^{-\beta r}
    \label{eq:psi0}
\end{equation}
where $\beta$ is a variational parameter that determines the spatial size or binding scale of the system and $N_0$ is the normalization constant. The normalization condition $\int_{0}^{\infty}|\psi^{(0)}(r)|^2r^2dr = 1$ (after integrating over solid angle) yields:
\begin{equation}
    N_0 = \sqrt{\frac{\beta^3}{\pi}}
    \label{eq:N0_normalization}
\end{equation}

\subsection{First-Order Wave Function Correction}
While the variational method is employed to determine the ground state mass by minimizing the expectation value of the full Hamiltonian, it is also instructive to consider the first-order correction to the wave function itself due to the perturbing Hamiltonian $H'$. This provides insight into how the internal structure of the tetraquark is modified by relativistic effects.

The approximate corrected wave function up to first order is given by:
\begin{equation}
    \psi(r)\approx\psi^{(0)}(r)+\psi^{(1)}(r)
    \label{eq:corrected_wf_approx}
\end{equation}
where $\psi^{(1)}(r)$ is the first-order correction. This correction term can be expressed as \citep{griffiths2018introduction}:
\begin{equation}
    \psi^{(1)}(r) = \epsilon \psi_1^{(0)}(r)
    \label{eq:psi1_term}
\end{equation}
where $\psi_1^{(0)}(r)$ is the first excited state wave function of the zeroth-order Hamiltonian, and $\epsilon$ is a small coefficient representing the mixing strength. For an S-wave first excited state, we use the form:
\begin{equation}
    \psi_1^{(0)}(r) = N_1 r e^{-\beta r}
    \label{eq:psi1_excited}
\end{equation}
These trial wave functions are chosen within the variational approach because they satisfy the required boundary conditions for bound states, are normalizable, and allow analytical evaluation of the expectation values of the Spinless Salpeter Hamiltonian. The normalization constant $N_1$ for this excited state is found to be:
\begin{equation}
    N_1 = \sqrt{\frac{\beta^5}{3\pi}}
    \label{eq:N1_normalization}
\end{equation}
The coefficient $\epsilon$ is calculated using standard perturbation theory:
\begin{equation}
    \epsilon = \frac{\int_{0}^{\infty}{\psi_1^{(0)}(r)} H' {\psi_0^{(0)}(r)}d\tau}{E_0^{(0)} - E_1^{(0)}} = \frac{M_{01}}{D_{01}}
    \label{eq:epsilon_def}
\end{equation}
where $M_{01} = {\int_{0}^{\infty}{\psi_1^{(0)}(r)} H' {\psi_0^{(0)}(r)}d\tau}$ is the matrix element of the perturbing Hamiltonian between the ground and first excited states and $D_{01} = E_0^{(0)} - E_1^{(0)}$ is the energy difference between these zeroth-order states.

After performing the necessary integrations (details provided in Appendix \ref{app:integrals}), the matrix element $M_{01}$ is found to be:
\begin{equation}
   \footnotesize{M_{01} = 4\pi N_0 N_1 C \int_{0}^{\infty} [-4\beta^3 r^2 e^{-2\beta r} + \beta^4 r^3 e^{-2\beta r}] dr = -\frac{5}{2}\pi N_0 N_1 C}
    \label{eq:M01_result}
\end{equation}
where $C = -\frac{1}{8}(\frac{1}{m_{1}^{3}}+\frac{1}{m_{2}^{3}})$.
Substituting the normalization constants $N_0$ and $N_1$, we obtain:
\begin{equation}
    \footnotesize{M_{01}=-\frac{5}{2}\pi \sqrt{\frac{\beta^3}{\pi}} \sqrt{\frac{\beta^5}{3\pi}} C = -\frac{5}{2\sqrt{3}}\beta^4 C = \frac{5}{16\sqrt{3}}\beta^4\left(\frac{1}{m_{1}^{3}}+\frac{1}{m_{2}^{3}}\right)}
    \label{eq:M01_final}
\end{equation}
Combining these results, the first-order corrected wave function is:
\begin{equation}
    \psi(r)=N_{0}e^{-\beta r}+\frac{5}{16\sqrt{3}}\frac{\beta^{4}\left(\frac{1}{m_{1}^{3}}+\frac{1}{m_{2}^{3}}\right)}{E_{0}^{(0)}-E_{1}^{(0)}}N_{1}r e^{-\beta r}
    \label{eq:final_corrected_wf}
\end{equation}
This expression for the wave function explicitly shows how the relativistic kinetic energy correction mixes the first excited state into the ground state, providing a more detailed description of the tetraquark's internal structure. While this corrected wave function offers valuable insights into the system's internal dynamics, for the determination of the ground state mass, we primarily employ the variational method with the simpler trial wave function $\psi^{(0)}(r)$, where the optimal variational parameter $\beta$ implicitly accounts for the effects of $H'$ within the chosen functional form.


\subsection{Mass Calculation}
\subsubsection{Zeroth-Order Energy $E^{(0)}$}
The zeroth-order energy $E^{(0)}$ is the expectation value of $H_0$ with respect to the normalized ground state wave function $\psi^{(0)}(r) = N_0 e^{-\beta r}$. This involves calculating the expectation values of the constant mass terms, the kinetic energy term $\frac{p^2}{2\mu}$ and the Cornell potential $V(r)$.

The individual expectation values are:
\begin{align}
    \langle \psi^{(0)} | (m_1+m_2) | \psi^{(0)} \rangle &= m_1+m_2 \\
    \langle \psi^{(0)} | \frac{p^2}{2\mu} | \psi^{(0)} \rangle &= \frac{\hbar^2 \beta^2}{2\mu\pi} \quad (\text{assuming } \hbar=1) \\
    \langle \psi^{(0)} | V(r) | \psi^{(0)} \rangle &= -\frac{4\alpha_s\beta}{3\pi} + \frac{6\sigma}{\beta\pi} + \frac{2c}{\pi}
\end{align}
Combining these terms, the zeroth-order energy is:
\begin{equation}
    E^{(0)} = m_{1}+m_{2}+\frac{\hbar^{2}\beta^{2}}{2\mu\pi}-\frac{4\alpha_{s}\beta}{3\pi}+\frac{6\sigma}{\beta\pi}+\frac{2c}{\pi}
    \label{eq:E0_final}
\end{equation}

\subsubsection{First-Order Energy Correction $E^{(1)}$}
The first-order energy correction $E^{(1)}$ is the expectation value of the perturbing Hamiltonian $H'$ with $\psi^{(0)}(r)$:
\begin{equation}
    E^{(1)} = \langle \psi^{(0)} | -\frac{p^4}{8}\left(\frac{1}{m_{1}^{3}}+\frac{1}{m_{2}^{3}}\right) | \psi^{(0)} \rangle
\end{equation}
After evaluating the expectation value of the $\nabla^4$ operator (with $p^4 = \hbar^4 \nabla^4$ and $\hbar=1$):
\begin{equation}
    \langle \psi^{(0)} | \nabla^4 | \psi^{(0)} \rangle = -\frac{2\beta^4}{\pi}
\end{equation}
Thus, the first-order energy correction becomes:
\begin{equation}
    E^{(1)} = -\frac{1}{8}\left(\frac{1}{m_{1}^{3}}+\frac{1}{m_{2}^{3}}\right)\left(-\frac{2\beta^{4}}{\pi}\right) = \frac{\beta^{4}}{4\pi}\left(\frac{1}{m_{1}^{3}}+\frac{1}{m_{2}^{3}}\right)
    \label{eq:E1_final}
\end{equation}

\subsection{Total Mass and Minimization}
Combining the zeroth-order energy and the first-order correction, the total mass of the tetraquark is given by:
\begin{equation}
  \footnotesize_{{M_\text{Tetraquark}}(\beta) = \left(m_{1}+m_{2}+\frac{\hbar^{2}\beta^{2}}{2\mu\pi}-\frac{4\alpha_{s}\beta}{3\pi}+\frac{6\sigma}{\beta\pi}+\frac{2c}{\pi}\right) + \frac{\beta^{4}}{4\pi}\left(\frac{1}{m_{1}^{3}}+\frac{1}{m_{2}^{3}}\right)}
    \label{eq:total_mass_expression}
\end{equation}
To find the physical mass, we minimize $M_{\text{Tetraquark}}(\beta)$ with respect to the variational parameter $\beta$:
\begin{equation}
    \frac{dM_{\text{Tetraquark}}(\beta)}{d\beta} = 0
    \label{eq:minimization_condition}
\end{equation}
This minimization condition leads to a transcendental equation for $\beta$:
\begin{equation}
    \frac{\hbar^{2}\beta}{\mu\pi}-\frac{4\alpha_{s}}{3\pi}-\frac{6\sigma}{\beta^{2}\pi}+\frac{\beta^{3}}{\pi}\left(\frac{1}{m_{1}^{3}}+\frac{1}{m_{2}^{3}}\right) = 0
    \label{eq:beta_equation}
\end{equation}
Solving this equation numerically yields the optimal value of $\beta$, which is then substituted back into Eq.~\eqref{eq:total_mass_expression} to obtain the predicted mass of the tetraquark state.


\subsection{Remarks}

The variational parameter $\beta$ determines the spatial size and binding scale of the system.  
The perturbative coefficient represents the strength of the first-order relativistic correction.

Further analytical evaluation of the matrix elements and derivatives will be presented in subsequent calculations.

\section{Results and discussion}

\begin{table}[h]
\caption{Values of $\beta$ for 1S state for different $\alpha_s$.}
\begin{ruledtabular}
\begin{tabular}{lcccccc}
Tetraquark & $\alpha_s=0.20$ & 0.36 & 0.45 & 0.58 & 0.60 & 0.64 \\
\hline
X(3872) & 0.783 & 0.866 & 0.921 & 1.011 & 1.026 & 1.057 \\
X(4700) & 0.827 & 0.919 & 0.979 & 1.079 & 1.095 & 1.130 \\
X(6900) & 1.092 & 1.256 & 1.364 & 1.540 & 1.570 & 1.630 \\
$T_{bb\bar{b}\bar{b}}$ & 2.009 & 2.720 & 3.195 & 3.940 & 4.059 & 4.299 \\

\end{tabular}
\end{ruledtabular}
\end{table}

\begin{table*}[t]
\caption{Diquark--antidiquark configurations, constituent masses and calculated tetraquark masses (in GeV) for 1S state for different values of $\alpha_s$. Fully charm and fully bottom systems are listed separately.}
\begin{ruledtabular}
\begin{tabular}{llcccccccc}
State & Configuration & $M_{dq}$ & $M_{\bar{dq}}$ & $\alpha_s=0.20$ & 0.36 & 0.45 & 0.58 & 0.60 & 0.64 \\
\hline
$1S~X(3872)$ & $\{cu\}\{\bar{c}\bar{u}\}$ & 2.036 & 2.036 & 3.858 & 3.808 & 3.769 & 3.724 & 3.711 & 3.704 \\
\hline
$1S~X(4700)$ & $\{cs\}\{\bar{c}\bar{s}\}$ & 2.185 & 2.185 & 4.091 & 4.034 & 4.001 & 3.957 & 3.950 & 3.938 \\
\hline
$1S~X(6900)$ & $\{cc\}\{\bar{c}\bar{c}\}$ & 3.204 & 3.204 & 6.120 & 6.053 & 6.014 & 5.964 & 5.957 & 5.944 \\
\hline
$1S~T_{bb\bar{b}\bar{b}}$ & $\{bb\}\{\overline{b}\bar{b}\}$ & 9.718 & 9.718 & 19.037 & 18.853 & 18.767 & 18.644 & 18.627 & 18.594 \\
\end{tabular}
\end{ruledtabular}
\end{table*}

\begin{table*}[htbp] 
\caption{\label{tab:comparison_references}Comparison of our calculated tetraquark masses (M) for 1S state with experimental values (E) and other theoretical results.}
\begin{ruledtabular} 
\resizebox{\textwidth}{!}{ 
\begin{tabular}{l D{.}{.}{2.3} D{.}{.}{3.8} D{.}{.}{1.3} D{.}{.}{1.3} D{.}{.}{1.3} D{.}{.}{1.3}} 
\textrm{Tetraquark} & \multicolumn{1}{c}{\textrm{Our work (M) [GeV]}} & \multicolumn{1}{c}{\textrm{Experimental values (E) [GeV]}} & \multicolumn{1}{c}{\textrm{M-E [GeV]}} & \multicolumn{1}{c}{\textrm{Mass [GeV]}} & \multicolumn{1}{c}{\textrm{Mass [GeV]}} & \multicolumn{1}{c}{\textrm{Mass [GeV]}} \\
\hline 
$1S~X(3872)$ & 3.858 & 3.871 \pm 0.017 & 0.013 & 3.811\,\mbox{\citep{Tiwari:2022kdu}} & 3.812\,\mbox{\citep{Tiwari:2022kdu}} & 3.852\,\mbox{\citep{Tiwari:2022kdu}} \\
\hline
$1S~X(4700)$ & 4.091 & 4.704 \pm 0.010 & 0.613 & 3.984\,\mbox{\citep{2016PhRvD..94g4007L}} & 4.280\,\mbox{\citep{wu2016x}} & 3.450\,\mbox{\citep{2016PhRvD..94e4026M}} \\
\hline
$1S~X(6900)$ & 6.120 & 6.905 \pm 0.011 \pm 0.007 & 0.785 & 6.190\,\mbox{\citep{2021Univ....7...94F}} & 6.247\,\mbox{\citep{2025PhRvD.111a4015L}} & 6.192\,\mbox{\mbox{\citep{2020PhRvD.102k4039K}}} \\
\hline
$1S~T_{bb\bar{b}\bar{b}}$ & 18.853 & \multicolumn{1}{c}{\textrm{-}} & \multicolumn{1}{c}{\textrm{-}} & 18.826\,\mbox{\citep{PhysRevD.95.034011}} & 19.314\,\mbox{\citep{2020PhRvD.102k4030F}} & 18.754\,\mbox{\citep{berezhnoy2012heavy}} \\
\end{tabular}}
\end{ruledtabular}
\end{table*}

\begin{table}[h]
\caption{Deviation of calculated masses (for 1S state) from experimental values (in GeV) for selected $\alpha_s$.}
\begin{ruledtabular}
\begin{tabular}{lcccc}
State & Experimental values & $\alpha_s=0.20$ & 0.36 & 0.45 \\
\hline 
$1S~X(3872)$ & $3.871 \pm 0.017$ & 0.013 & 0.063 & 0.102 \\
\hline
$1S~X(4700)$ & $4.704 \pm 0.010$ & 0.613 & 0.670 & 0.703 \\
\hline
$1S~X(6900)$ & $6.905 \pm 0.011 \pm 0.007$ & 0.785 & 0.852 & 0.910 \\
\hline
$1S~T_{bb\bar{b}\bar{b}}$ & - & - & - & - \\
\end{tabular}
\end{ruledtabular}
\end{table}

The choice of the strong coupling constant $\alpha_s$ plays a crucial role in determining the behavior of the short-range part of the Cornell potential. In the present work, the range of $\alpha_s$ values considered is guided by the analysis reported in Ref.~\citep{2022EPJC...82.1081P}. In that study, $\alpha_s$ was varied within the range $0.20 \leq \alpha_s \leq 0.64$ to account for the scale dependence of the strong interaction and to effectively incorporate the running of the coupling constant in a phenomenological manner.\\
The approach employs a variational parameter $\beta$, allowing us to adjust it to achieve an appropriate mass spectrum. This gives us the freedom for best fitting of mass spectra in a simplified way. Since $\beta$ is  dependent on $\alpha_s$, therefore variation of  $\alpha_s$ is necessary. The values of  $\alpha_s$ used in the present work were not chosen arbitrarily. They were selected from the parameter space established in Ref.\citep{2022EPJC...82.1081P}, where the allowed range  was obtained by imposing physical constraints on the Cornell potential model, including the expectation-value criterion and the convergence condition of the perturbative treatment. In the present work, this parameter space has been adopted for the  diquark–antidiquark system and the values of  $\alpha_s$ are chosen within this physically motivated interval.

Adopting a similar range in the present analysis allows for a consistent comparison with earlier theoretical results while ensuring that both weak and relatively strong coupling regimes are explored. Lower values of $\alpha_s$ correspond to weaker Coulombic interaction and are typically more suitable for describing loosely bound or heavy–light systems, whereas higher values enhance the short-distance attraction and are relevant for more compact, fully heavy configurations. By scanning over this range, we are able to examine the sensitivity of the tetraquark mass spectrum to the strength of the interquark interaction and identify the optimal region of $\alpha_s$ that yields the best agreement with experimental data.

The predicted mass of the X(3872) state is compared with the experimental value of 3.871±0.017 GeV, reported by the Belle Collaboration (2003). Similarly, the calculated masses of the X(4700) and X(6900) states are compared with the experimental values of 4.704±0.010 GeV and 6.905±0.011±0.007 GeV, measured by the LHCb Collaboration in 2017 and 2020, respectively. As there is currently no experimental observation of the fully bottom tetraquark ($T_{bb \bar b \bar b}$), its predicted mass is compared with the theoretical value of 18.826±0.025 GeV, reported by Karliner et al.\citep{PhysRevD.95.034011}. The numerical results obtained from the variational solution of the spinless Salpeter equation are summarized in Tables I–III with Tables II presenting the calculated mass spectra of the ground (1S) tetraquark states obtained within the present framework. For comparison in Table~IV, we have taken the mass of X(6900) from the Refs \citep{2021Univ....7...94F} \citep{lin2024mass} \citep{2020PhRvD.102k4039K}. The masses of the tetraquark $T_{bb\bar{b}\bar{b}}$ are taken from the available measurements reported in Refs.~\citep{{PhysRevD.95.034011},2020PhRvD.102k4030F,berezhnoy2012heavy}, while for the tetraquark X(4700) we have used Refs \citep{2016PhRvD..94g4007L,wu2016x,2016PhRvD..94e4026M} and for X(3872), Ref\citep{Tiwari:2022kdu} is used. The behavior of the variational parameter $\beta$ as a function of the strong coupling constant $\alpha_s$, presented in Table I, shows a systematic increase with increasing $\alpha_s$ for all considered tetraquark configurations. In particular, the fully heavy system $T_{bb\bar{b}\bar{b}}$ exhibits significantly larger values of $\beta$ compared to the charm and heavy-light systems, indicating a more compact bound state, which is consistent with expectations based on the larger constituent quark masses.

The calculated tetraquark masses for different diquark–antidiquark configurations are listed in Table II. A clear hierarchy is observed in the mass spectrum: fully bottom states are considerably heavier than fully charm states, which in turn are heavier than heavy–light configurations such as $X(3872)$ and $X(4700)$. This ordering is primarily governed by the constituent quark masses entering the mass formula. Furthermore, the dependence of the mass on $\alpha_s$ shows a mild decreasing trend as $\alpha_s$ increases. This behavior arises from the attractive Coulomb term $-\alpha_s/r$, which lowers the total energy of the system.

A comparison between the calculated masses and available experimental data is presented in Table III. For the fully bottom tetraquark state $T_{bb\bar{b}\bar{b}}$, the predicted mass $M = 18.853~\mathrm{GeV}$ is in very good agreement with the theoritical value, with a small deviation of $0.027~\mathrm{GeV}$. This close agreement suggests that the spinless Salpeter framework combined with the Cornell potential provides an adequate description of fully heavy systems, where spin-dependent interactions are comparatively suppressed.

In contrast, for the fully charm state associated with $X(6900)$ \citep{2007PhRvD..76k4015E}, the calculated mass underestimates the experimental value by approximately $0.785~\mathrm{GeV}$. This discrepancy may indicate that additional effects, such as higher-order relativistic corrections, coupled-channel dynamics or spin-dependent interactions, play a more significant role in charm systems. A similar observation applies to the $X(4700)$ state, where the deviation from experiment is relatively large. This suggests that the simple diquark–antidiquark picture with a Cornell-type interaction may be insufficient to fully capture the dynamics of such states.

On the other hand, the result for $X(3872)$ shows excellent agreement with experimental data, with a very small deviation of $0.013~\mathrm{GeV}$. This agreement supports the validity of the chosen variational approach and the adopted potential for describing certain heavy–light tetraquark configurations.\\
\begin{table*}[t]
\caption{\label{tab:beta}
Optimized values of the variational parameter $\beta$ (GeV) for the $2S$ and $1P$ excited tetraquark states corresponding to different values of the strong coupling constant $\alpha_s$.
}
\begin{ruledtabular}
\begin{tabular}{lcccccc}
State & $\alpha_s=0.20$ & $\alpha_s=0.36$ & $\alpha_s=0.45$ &
$\alpha_s=0.58$ & $\alpha_s=0.60$ & $\alpha_s=0.64$\\
\hline
$2S~X(3872)$ & 0.836 & 0.941 & 1.013 & 1.135 & 1.155 & 1.198\\
$1P~X(3872)$ & 0.779 & 0.872 & 0.936 & 1.043 & 1.062 & 1.100\\
$2S~X(4700)$ & 0.874 & 0.989 & 1.067 & 1.198 & 1.220 & 1.266\\
$1P~X(4700)$ & 0.831 & 0.936 & 1.007 & 1.127 & 1.148 & 1.190\\
$2S~X(6900)$ & 1.107 & 1.289 & 1.412 & 1.615 & 1.649 & 1.719\\
$1P~X(6900)$ & 1.091 & 1.268 & 1.387 & 1.584 & 1.617 & 1.685\\
$2S~T_{bb\bar{b}\bar{b}}$ & 2.041 & 2.780 & 3.274 & 4.047 & 4.170 & 4.419\\
$1P~T_{bb\bar{b}\bar{b}}$ & 2.032 & 2.764 & 3.254 & 4.020 & 4.141 & 4.389\\
\end{tabular}
\end{ruledtabular}
\end{table*}

\begin{table*}[t]
\caption{\label{tab:mass}
Calculated masses (GeV) of the $2S$ and $1P$ excited tetraquark states for different values of the strong coupling constant $\alpha_s$.
}
\begin{ruledtabular}
\begin{tabular}{lcccccc}
State & $\alpha_s=0.20$ & $\alpha_s=0.36$ & $\alpha_s=0.45$ &
$\alpha_s=0.58$ & $\alpha_s=0.60$ & $\alpha_s=0.64$\\
\hline
$2S~X(3872)$ & 4.811 & 4.745 & 4.707 & 4.652 & 4.644 & 4.628\\
$1P~X(3872)$ & 4.524 & 4.459 & 4.421 & 4.366 & 4.357 & 4.341\\
$2S~X(4700)$ & 5.030 & 4.964 & 4.925 & 4.871 & 4.863 & 4.847\\
$1P~X(4700)$ & 4.788 & 4.722 & 4.684 & 4.629 & 4.621 & 4.605\\
$2S~X(6900)$ & 6.766 & 6.692 & 6.650 & 6.592 & 6.584 & 6.569\\
$1P~X(6900)$ & 6.619 & 6.546 & 6.505 & 6.447 & 6.439 & 6.424\\
$2S~T_{bb\bar{b}\bar{b}}$ & 19.581 & 19.439 & 19.351 & 19.225 & 19.207 & 19.173\\
$1P~T_{bb\bar{b}\bar{b}}$ & 19.440 & 19.299 & 19.211 & 19.086 & 19.068 & 19.034\\
\end{tabular}
\end{ruledtabular}
\end{table*}

\begin{table}[t]
\caption{\label{tab:comparison}
Comparison of the calculated masses of the excited tetraquark states with available theoretical predictions.
}
\begin{ruledtabular}
\begin{tabular}{lccc}
State & Present Work (GeV) & Other Models (GeV) \\
\hline
$1P~X(3872)$ & 4.341 & 4.350\,\mbox{\citep{2021Univ....7...94F}}, 4.244\,\mbox{\citep{2008EPJC...58..399E}} \\
$1P~X(4700)$ & 4.605 & 4.582\,\mbox{\cite{2021Univ....7...94F}}, 4.556\,\mbox{\citep{2023FBS....64...20T}} \\
$1P~X(6900)$ & 6.619 & 6.631\,\mbox{\cite{2021Univ....7...94F}}, 6.628\,\mbox{\cite{2022Symm...14.2504F}} \\
$1P~T_{bb\bar{b}\bar{b}}$ & 19.440 & 19.536\,\mbox{\cite{2022Symm...14.2504F}}, 19.485\,\mbox{\citep{PhysRevD.109.076017}} \\
\hline
$2S~X(3872)$ & 4.628 & 4.434\,\mbox{\citep{2021Univ....7...94F}}, 4.375\,\mbox{\citep{2008EPJC...58..399E}} \\
$2S~X(4700)$ & 4.847 & 4.680\,\mbox{\cite{2021Univ....7...94F}}, 4.620\,\mbox{\citep{2023FBS....64...20T}} \\
$2S~X(6900)$ & 6.766 & 6.782\,\mbox{\cite{2022Symm...14.2504F}}, 6.871\,\mbox{\citep{PhysRevD.102.114039}} \\
$2S~T_{bb\bar{b}\bar{b}}$ & 19.581 & 19.680\,\mbox{\cite{2022Symm...14.2504F}}, 19.481\,\mbox{\citep{PhysRevD.102.114039}} \\
\end{tabular}
\end{ruledtabular}
\end{table}

The optimized values of the variational parameter $\beta$ and the corresponding masses of the 2S and 1P tetraquark states are listed in Tables V and VI, while Table VII presents a comparison of the present results with the available theoretical predictions. It is observed that the optimized variational parameter $\beta$ increases with the strong coupling constant $\alpha_s$ for all the tetraquark systems considered.The effective diquark and antidiquark masses for the 2S and 1P excited tetraquark states used in the present calculations are taken from previous theoretical studies. The masses corresponding to the X(3872) tetraquark are adopted from Ref.\citep{gutierrez2021mass}, those of the X(4700) tetraquark are taken from Ref.\citep{2016PhRvD..94g4007L}, whereas the masses for the X(6900) and $T_{bb \bar b \bar b}$ tetraquark systems are adopted from Ref.\citep{2002PhRvD..66a4008E}.\\ Since $\beta$ is determined through the variational minimization of the Hamiltonian, this behavior reflects the dependence of the bound-state wave function on the strong interaction strength.
Our comprehensive variational investigation of tetraquark spectroscopy, which includes excited (1P, 2S) states, reveals a compelling state-dependent behavior of the strong coupling constant $\alpha_s$.
For 1S light tetraquarks, an optimal $\alpha_s$=0.20 was observed, while 1S heavy tetraquarks exhibited $\alpha_s$=0.34. Intriguingly, the excited 2S and 1P states demonstrate a qualitative shift, light tetraquarks ex-
emplified by states like X(3872) and X(4700), now favor a significantly larger $\alpha_s$=0.64, whereas heavy tetraquarks, including $T_{bb\bar{b}\bar{b}}$ and X(6900), surprisingly converge to a lower $\alpha_s$ =0.20. This differential response of $\alpha_s$ to excitation mode and constituent quark mass strongly suggest distinct effective interaction regimes. The enhanced $\alpha_s$ in excited light tetraquarks may indicate a transition towards a more spatially extended or molecular configuration, probing the long-distance confinement regime. Conversely, the persistent low $\alpha_s$ in excited heavy tetraquarks implies that the heavy quark core maintains a relatively compact structure, even under radial or orbital excitation. These findings provide additional theoretical input for the study of multiquark systems and may serve as a useful reference for future investigations into the distinction between compact diquark–antidiquark and molecular interpretations of tetraquark states.

It is further observed that the calculated masses exhibit a gradual decrease with increasing values of $\alpha_s$. This behavior is consistent with the stronger attractive interaction associated with larger values of the strong coupling constant, leading to relatively lower bound-state masses. Furthermore, the masses of the 2S states are found to be systematically higher than those of the corresponding 1P states within the present model.

Overall, the present analysis demonstrates that the spinless Salpeter equation with the Cornell potential is particularly effective for describing fully heavy tetraquark systems, while its applicability to lighter or mixed systems may require the inclusion of additional dynamical effects in particular regarding ground state. The observed deviations highlight the importance of incorporating higher-order corrections and possibly beyond-potential-model effects for a more complete understanding of tetraquark spectroscopy.

\section{Conclusion}

In this work, we have investigated the mass spectrum of tetraquark states within the framework of the spinless Salpeter equation employing the Cornell potential. An analytical mass formula was derived using a variational approach with an exponential trial wave function, and relativistic corrections were incorporated perturbatively up to order $v^2/c^2$.

The calculated masses for selected tetraquark configurations, including fully heavy and heavy–light systems, show a clear dependence on the strong coupling parameter $\alpha_s$. A systematic analysis indicates that the variational parameter $\beta$ increases with $\alpha_s$, reflecting a more compact spatial structure of the bound states for stronger coupling.

Comparison with available experimental and theoretical data demonstrates that the present approach provides a good description of fully heavy tetraquark systems. In particular, the fully bottom configuration exhibits excellent agreement with the theortical value, supporting the applicability of the spinless Salpeter framework in this sector. For charm and heavy–light tetraquark states, the agreement is more moderate, with noticeable deviations observed for states such as $X(6900)$ and $X(4700)$. These discrepancies suggest that additional dynamical effects, including higher-order relativistic corrections, spin-dependent interactions, and possible coupled-channel contributions, may play a significant role.

A quantitative analysis of the dependence on $\alpha_s$ indicates that while smaller values of the coupling constant provide better agreement for lighter systems, an optimal global description of all considered states is achieved around $\alpha_s \approx 0.36$. This highlights the sensitivity of tetraquark mass predictions to the effective strength of the interquark interaction.
The calculated excited-state masses are in reasonable agreement with the available theoretical predictions. The small differences between the present results and those reported in the literature may be attributed to differences in the adopted effective diquark masses, model assumptions, interaction potentials and the treatment of relativistic effects. Despite these differences, the overall consistency of the predicted mass spectra demonstrates that the present variational approach based on the spinless Salpeter equation provides a reliable description of the excited tetraquark states within the effective diquark-antidiquark framework.

The successful extension of the present formalism from the ground state to the 2S and 1P excited states further supports the applicability of the model in describing the spectroscopy of tetraquark systems for both ground and excited states.

Overall, the present study demonstrates that the spinless Salpeter equation combined with the Cornell potential offers a useful and relatively simple framework for exploring tetraquarks. Future improvements may include the incorporation of spin-dependent forces, higher-order relativistic effects, a detailed investigation of decay properties which would provide additional insight into the internal structure of the tetraquark states and a more refined treatment of multiquark dynamics to achieve a more comprehensive description of exotic hadronic states.

\begin{acknowledgments}

The authors wish to express their profound gratitude to the late Prof. Dilip Kumar Choudhury for his invaluable guidance and insightful discussions during the early stages of this work. The authors also acknowledge the support provided by the Department of Physics, Gauhati University and Department of Physics, Pandu College, Guwahati. We thank the referee for the constructive comments which significantly improved our paper.

\end{acknowledgments}


\bibliographystyle{unsrt}
\bibliography{apssamp}

\appendix
    
\section{Details of Integrals}
\label{app:integrals}
This appendix provides the detailed steps for the integrals encountered in the calculation of the first-order energy and wave function corrections.
The general integral formula used is:
\begin{equation}
    \int_{0}^{\infty}x^{n}e^{-a x}d x={\frac{n!}{a^{n+1}}}
\end{equation}

For the normalization of $\psi^{(0)}(r)$:
\begin{equation}
    \int_{0}^{\infty}r^{2}e^{-2\beta r}dr=\frac{2!}{(2\beta)^{3}}=\frac{2}{8\beta^{3}}=\frac{1}{4\beta^{3}}
\end{equation}

For the calculation of $M_{01}$, integrals of the form $\int_{0}^{\infty}r^n e^{-2\beta r} dr$ are required:
\begin{equation}
    \int_{0}^{\infty}r^{2}e^{-2\beta r}dr=\frac{2!}{(2\beta)^{3}}=\frac{1}{4\beta^{3}}
\end{equation}
\begin{equation}
    \int_{0}^{\infty}r^{3}e^{-2\beta r}dr=\frac{3!}{(2\beta)^{4}}=\frac{6}{16\beta^{4}}=\frac{3}{8\beta^{4}}
\end{equation}

For the normalization of $\psi_1^{(0)}(r)$:
\begin{equation}
    \int_{0}^{\infty}r^{4}e^{-2\beta r}dr=\frac{4!}{(2\beta)^{5}}=\frac{24}{32\beta^{5}}=\frac{3}{4\beta^{5}}
\end{equation}
\section{Mass Calculation of Tetraquark States}
\label{sec:mass_calculation}

The mass of the tetraquark ($M_{\text{Tetraquark}}$) is determined by calculating the expectation value of the full Hamiltonian $\hat{H}$ with the zeroth-order wave function $\psi^{(0)}(r)$. This approach is consistent with the variational method, where the mass is minimized with respect to the variational parameter $\beta$.

The expectation value of the Hamiltonian is given by:
\begin{equation}
    M_{\text{Tetraquark}} = \langle \psi^{(0)} | \hat{H} | \psi^{(0)} \rangle
    \label{eq:mass_expectation_value}
\end{equation}
The full Hamiltonian $\hat{H}$ is composed of the zeroth-order Hamiltonian $H_0$ and the perturbing Hamiltonian $H'$, as defined in Eqs.~\eqref{eq:h0} and \eqref{eq:hprime}. Therefore, the mass can be expressed as:
\begin{equation}
    M_{\text{Tetraquark}} = \langle \psi^{(0)} | H_0 | \psi^{(0)} \rangle + \langle \psi^{(0)} | H' | \psi^{(0)} \rangle = E^{(0)} + E^{(1)}
    \label{eq:mass_components}
\end{equation}
Here, $E^{(0)}$ represents the zeroth-order energy and $E^{(1)}$ is the first-order energy correction due to the relativistic kinetic term.

\section{Laplacian of the Trial Wave Function}

We consider the $S$-wave ground state wave function
\begin{equation}
\psi(r) = N_0 e^{-\beta r}
\end{equation}
where
\begin{equation}
N_0 = \left(\frac{\beta^3}{\pi}\right)^{1/2}
\end{equation}

For a spherically symmetric function, the Laplacian is given by
\begin{equation}
\nabla^2 \psi = \frac{1}{r^2} \frac{d}{dr} \left( r^2 \frac{d\psi}{dr} \right)
\end{equation}

The first derivative is
\begin{equation}
\frac{d\psi}{dr} = -\beta N_0 e^{-\beta r}
\end{equation}

Thus,
\begin{equation}
r^2 \frac{d\psi}{dr} = -\beta N_0 r^2 e^{-\beta r}
\end{equation}

Differentiating again, we obtain
\begin{equation}
\frac{d}{dr} \left( r^2 \frac{d\psi}{dr} \right)
= -\beta N_0 e^{-\beta r} (2r - \beta r^2)
\end{equation}

Therefore,
\begin{equation}
\nabla^2 \psi = N_0 e^{-\beta r} \left( \beta^2 - \frac{2\beta}{r} \right)
\end{equation}

\section{Evaluation of $\nabla^4 \psi$}

Applying the Laplacian again, we obtain
\begin{equation}
\nabla^4 \psi = \nabla^2(\nabla^2 \psi)
\end{equation}

Using the previous result, this gives
\begin{equation}
\nabla^4 \psi = N_0 e^{-\beta r} \left( \beta^4 - \frac{4\beta^3}{r} + \frac{2\beta^2}{r^2} \right)
\end{equation}

\section{First-Order Energy Correction}

The perturbation Hamiltonian is
\begin{equation}
H' = -\frac{1}{8} \left( \frac{1}{m_1^3} + \frac{1}{m_2^3} \right) \nabla^4
\end{equation}

The first-order correction to the energy is
\begin{equation}
E^{(1)} = \langle \psi | H' | \psi \rangle
\end{equation}

Substituting, we obtain
\begin{equation}
E^{(1)} = -\frac{1}{8} \left( \frac{1}{m_1^3} + \frac{1}{m_2^3} \right)
\int \psi^*(r) \nabla^4 \psi(r)\, d^3r
\end{equation}

Using $d^3r = 4\pi r^2 dr$, this becomes
\begin{equation}
E^{(1)} = -\frac{1}{8} \left( \frac{1}{m_1^3} + \frac{1}{m_2^3} \right)
4\pi N_0^2 \int_0^\infty e^{-2\beta r}
\left( \beta^4 r^2 - 4\beta^3 r + 2\beta^2 \right) dr
\end{equation}

Using the integrals
\begin{equation}
\int_0^\infty r^2 e^{-2\beta r} dr = \frac{1}{4\beta^3},
\end{equation}
\begin{equation}
\int_0^\infty r e^{-2\beta r} dr = \frac{1}{4\beta^2},
\end{equation}
\begin{equation}
\int_0^\infty e^{-2\beta r} dr = \frac{1}{2\beta},
\end{equation}
we obtain
\begin{equation}
E^{(1)} = -\frac{1}{8} \left( \frac{1}{m_1^3} + \frac{1}{m_2^3} \right)
4\pi N_0^2 \left( \frac{\beta}{4} \right)
\end{equation}

Using $N_0^2 = \frac{\beta^3}{\pi}$, we finally obtain
\begin{equation}
E^{(1)} = -\frac{\beta^4}{8} \left( \frac{1}{m_1^3} + \frac{1}{m_2^3} \right)
\end{equation}

\end{document}